\documentclass[12pt,a4paper]{article}
\usepackage{graphicx}
\usepackage{dcolumn}

\begin{document}
\title{Light-ion production in the interaction\\
of 96 MeV neutrons with silicon}
\author{ U. Tippawan$^{1,2}$, 
S. Pomp$^{1}$\footnote{Corresponding author, Tel. +46 18 471 6850, 
Fax. +46 18 471 3853, E-mail: stephan.pomp@tsl.uu.se}, 
A. Ata\c{c}$^{1}$, B. Bergenwall$^1$, J. Blomgren$^1$, \\
S. Dangtip$^{1,2}$, A. Hildebrand$^1$, C. Johansson$^1$, J. Klug$^1$, P. 
Mermod$^1$, \\
L. Nilsson$^{1,4}$, M. \"Osterlund$^1$, N. Olsson$^{1,3}$, K. Elmgren$^3$, \\
O. Jonsson$^4$, A.V. Prokofiev$^4$, P.-U. Renberg$^4$, P. Nadel-Turonski$^5$, \\
V. Corcalciuc$^6$, Y. Watanabe$^7$, A. Koning$^8$ \\
\\
\fontsize{9}{12}
  $^{1}$\textit{Department of Neutron Research, Uppsala University, Sweden} \\
\fontsize{9}{12}
  $^{2}$\textit{Fast Neutron Research Facility, Chiang Mai University, Thailand 
} \\
\fontsize{9}{12}
  $^{3}$\textit{Swedish Defence Research Agency, Stockholm, Sweden} \\
\fontsize{9}{12}
  $^{4}$\textit{The Svedberg Laboratory, Uppsala University, Sweden} \\
\fontsize{9}{12}
  $^{5}$\textit{Department of Radiation Sciences, Uppsala University, Sweden} \\
\fontsize{9}{12}
  $^{6}$\textit{Institute of Atomic Physics, Heavy Ion Department, Bucharest, 
Romania} \\
\fontsize{9}{12}
  $^{7}$\textit{Department of Advanced Energy Engineering Science, Kyushu 
University, Japan} \\
\fontsize{9}{12}
  $^{8}$\textit{Nuclear Research and Consultancy Group, Petten, The 
Netherlands}}
\date{}
\maketitle
\section*{Abstract}
Double-differential cross sections for light-ion (p, d, t, $^3$He and $\alpha$) 
production in 
silicon, induced by 96 MeV neutrons are reported. Energy spectra are measured at 
eight 
laboratory angles from $20^\circ$ to $160^\circ$ in steps of $20^\circ$. 
Procedures for data 
taking and data reduction are presented. Deduced energy-differential, 
angle-differential and 
production cross sections are reported. Experimental cross sections are compared 
to 
theoretical reaction model calculations and experimental data in the literature.
\\
PACS numbers: 25.40.-h, 25.40.Hs, 25.40.Kv, 28.20.-v



\section{Introduction}
\label{sec:Introduction}


In the last years, there has been an increasing request for experimental studies 
of fast-neutron 
induced reactions, especially at higher incident neutron energies. For basic 
physics, nucleon-induced 
reactions provide useful means to investigate nuclear structure, to characterize 
reaction mechanisms 
and to impose stringent constraints on nuclear model calculations. The silicon 
nucleus is sufficiently 
heavy for many of the statistical assumptions to hold (high density of excited 
states), yet not so 
heavy to give a strong suppression of charged particle emission due to Coulomb 
barrier effects. 
Therefore, nuclear reaction models for equilibrium and pre-­equilibrium decay 
can be tested and 
benchmarked. Experimental data in the literature at incident neutron energies 
from reaction thresholds 
up to 60 MeV~\cite{Bat99} and between 25 and 65 MeV~\cite{Ben02a} offer 
possibilities to test the 
predictions of reaction models.

In recent years, an increasing number of applications involving fast neutrons 
have been developed or 
are under consideration, e.g., radiation treatment of 
cancer~\cite{Ore98,Sch01,Lar95}, soft-error 
effects in computer memories~\cite{Single}, accelerator-driven transmutation of 
nuclear waste and 
energy production~\cite{High}, and determination of the response of neutron 
detectors~\cite{Cec79}. 
Silicon data are particularly important for detailed soft-error simulation in 
electronic devices~\cite
{Single,Cha99}.

In this paper, we present experimental double-differential cross sections 
(inclusive yields) for protons,
deuterons, tritons, $^3$He and alpha particles produced by 96 MeV neutrons 
incident on silicon. Measurements 
have been performed at the cyclotron of The Svedberg Laboratory (TSL), Uppsala, 
using the dedicated 
MEDLEY experimental setup~\cite{Dan00}. Spectra have been measured at 8 
laboratory angles, ranging from 
$20^\circ$ to $160^\circ$ in $20^\circ$ steps. Extrapolation procedures are used 
to obtain coverage of 
the full angular distribution and consequently energy-differential and 
production cross sections are 
deduced, the latter by integrating over energy and angle. The experimental data 
are compared to results 
of calculations with nuclear reaction codes and to existing experimental data.

The experimental methods are briefly discussed in sect.~\ref{sec:Experimental 
setup} and data reduction and 
correction procedures are presented in sect.~\ref{sec:Data reduction} 
and~\ref{sec:Corrections}, respectively. 
The theoretical framework is presented in sect.~\ref{sec:Theoretical models}. In 
sect.~\ref{sec:Results and discussion}, 
experimental results are reported and compared with theoretical and previous 
experimental data. Conclusions 
and an outlook are given in sect.~\ref{sec:Conclusions and outlook}. 


\section{Experimental setup and methods}
\label{sec:Experimental setup}


The neutron beam facility at TSL uses the $^7$Li(p,n)$^7$Be reaction (Q $= 
-1.64$ MeV) to produce a 
quasi-monoenergetic neutron beam~\cite{Klu01}. The lithium target was 26 mm in 
diameter and 8 mm thick 
in the present experiment and enriched to 99.98 \% in $^7$Li. The 98.5$\pm$0.3 
MeV protons from the 
cyclotron impinge on the lithium target, producing a full-energy peak of 
neutrons at 95.6$\pm$0.5 MeV 
with a width of 1.6 MeV (FWHM) and containing 40 \% of the neutrons, and an 
almost constant low-energy tail 
containing 60 \% of the neutrons. The neutron beam is shaped by a collimator 
system, and delivered to 
the experimental area. After passage of the target, the proton beam is deflected 
by two magnets into a 
well-shielded beam dump, where the beam current is integrated in a Faraday cup. 
The integrated charge 
serves as one neutron beam monitor. With a beam intensity of about 5 $\mu$A, the 
neutron flux at the target 
location is about 5$\cdot$10$^4$ neutrons/(s$\cdot$cm$^2$). The collimated 
neutron beam has a diameter of 
80 mm at the location of the target. A thin film breakdown counter 
(TFBC)~\cite{Smi95} installed after the 
reaction chamber is used as another beam monitor. The two beam monitor readings 
were in agreement during 
the measurements. 

The charged particles are detected by the MEDLEY setup~\cite{Dan00}. It consists 
of eight three-element 
telescopes mounted inside a 100 cm diameter evacuated reaction chamber. Each 
telescope consists of two 
fully depleted $\Delta E$ silicon surface barrier detectors and a CsI(Tl) 
crystal. The thickness of the 
first $\Delta E$ detector ($\Delta E_1$) is either 50 or 60 $\mu$m, while the 
second one ($\Delta E_2$) 
is either 400 or 500 $\mu$m. They are all 23.9 mm in diameter (nominal). The 
cylindrical CsI(Tl) crystal, 
50 mm long and 40 mm in diameter, serves as the $E$ detector. The back-end part 
of the crystal, 20 mm 
long, has a conical shape, tapered off to 18 mm diameter, to fit the size of a 
read-out diode. 

To obtain a well-defined acceptance, a plastic scintillator collimator is placed 
in front of each 
telescope. The active collimators have an opening of 19 mm diameter and a 
thickness of 1 mm. 

A Passivated Implanted Planar Silicon (PIPS) detector is used as an active 
target. It has a 32$\cdot$32 
mm$^2$ quadratic shape and a thickness of 303 $\mu$m. It is suspended in a thin 
aluminium frame using 
threads and small springs. The dimensions of the frame have been chosen in such 
a way that it does not 
interfere with the incident neutron beam. Besides the energy deposited by the 
detected light ion, the 
active target recorded the energy deposition due to other products, like 
recoils, of the same event. 
This information was, however, not used in the present analysis.

For absolute cross section normalization, a 25 mm diameter and 1.0 mm thick 
polyethylene (CH$_2$)$_n$ 
target is used. The $np$ cross section at $20^\circ$ laboratory angle provides 
the reference cross 
section~\cite{Rah01}.

Background is measured by removing the target from the neutron beam. It is 
dominated by protons 
produced by neutron beam interaction with the beam tube and reaction chamber 
material, especially at 
the entrance and exit of the reaction chamber and in the telescope housings. 
Therefore, the telescopes 
at $20^\circ$ and $160^\circ$ are most affected. Since the protons in the 
background originated not from 
the target but came from different directions, they can be misidentified leading 
to a large background 
even for the other particles. For the $160^\circ$ telescope, i.e., the worst 
case, the signal-to-background 
ratios are 2.5, 1 and 0.1 for protons, deuterons and tritons, respectively, 
whereas the corresponding 
numbers for the $40^\circ$ telescope, i.e., the best case, are 8, 12 and 5. In 
the case of $^3$He and 
alpha particles, the background is negligible.

The time-of-flight (TOF) obtained from the radio frequency of the cyclotron 
(stop signal for the TDC) and 
the timing signal from each of the eight telescopes (start signal), is measured 
for each charged-particle 
event.

The raw data are stored event by event for on-line monitoring and subsequent 
off-line analysis. 
Typical count rates for target-in and target-out runs were 10 and 2 Hz, 
respectively. The dead time 
of the system was typically 1-2  \% and never exceeding 10  \%.


\section{Data reduction procedures}
\label{sec:Data reduction}


\subsection{Particle identification and energy calibration}
\label{subsec:Particle identification}

The $\Delta E - E$ technique is used to identify light charged particles ranging 
from protons to lithium 
ions, which is illustrated in Fig.~\ref{fig:fig1}a. Good separation of all 
particles is obtained 
over their entire energy range. Since the energy resolution of each individual 
detector varies with 
the particle type, the particle identification cuts are defined to cover 
3$\sigma$, where $\sigma$ is the 
standard deviation of the energy resolution of each particle type. Typical 
energy resolutions of the thin 
$\Delta E$ detectors are between 40 and 80 keV, increasing with particle mass. 
The corresponding values 
are between 150 and 550 keV for the thick $\Delta E$ detectors and between 900 
and 1200 keV for the $E$ 
detectors. For energy depositions in the $E$ detector above 70 MeV in 
Fig.~\ref{fig:fig1}b, the two-dimensional 
cuts for protons, deuterons and tritons overlap slightly since the energy loss 
of the hydrogen isotopes 
in the $\Delta E_2$ detector is rather small. This ambiguity is resolved by a 
two-dimensional plot (inset 
of Fig.~\ref{fig:fig1}b) of the deviations of the $\Delta E_1$ and $\Delta E_2$ 
signals from tabulated energy 
loss values in silicon~\cite{Ziegler} (solid lines in Fig.~\ref{fig:fig1}a). 
Particle identification is done by 
cutting along the minimum contour line, and thus possible misidentification 
should even out. This technique is 
also used to improve the separation between $^3$He and alpha particles in some 
telescopes where the energy 
resolution is poor.

Energy calibration of all detectors is obtained from the data 
itself~\cite{Tip02}. Events in the 
$\Delta E - E$ bands are fitted with respect to the energy deposited in the 
three detectors (solid lines in 
Fig.~\ref{fig:fig1}). This energy is determined from the detector thicknesses 
and tabulated energy loss 
values in silicon~\cite{Ziegler}. The $\Delta E_1$ detectors are further 
calibrated and checked using a 
5.48 MeV alpha source. For the energy calibration of the CsI(Tl) detectors, two 
parameterizations of the 
light output versus energy of the detected particle~\cite{Dan00} are used, one 
for hydrogen isotopes and 
another one for helium isotopes. Supplementary calibration points are provided 
by the H(n,p) reaction, as 
well as transitions to the ground state and low-lying states in the 
$^{12}$C(n,d)$^{11}$B and $^{28}$Si
(n,d)$^{27}$Al reactions. The energy of each particle type is obtained by adding 
the energy deposited in 
each element of the telescope.

Low-energy charged particles are stopped in the $\Delta E_1$ detector leading to 
a low-energy cutoff for 
particle identification of about 3 MeV for hydrogen isotopes and about 8 MeV for 
helium isotopes 
(see Fig.~\ref{fig:fig1}a). The helium isotopes stopped in the $\Delta E_1$ 
detector are nevertheless 
analyzed and a remarkably low cutoff, about 4 MeV, can be achieved for the 
experimental alpha-particle 
spectra. These alpha-particle events could obviously not be separated from 
$^3$He events in the same energy 
region, but the yield of $^3$He is much smaller than the alpha-particle yield in 
the region just above 
8 MeV, where the particle identification works properly. That the relative yield 
of $^3$He is small is 
also supported by the theoretical calculations in the evaporation peak region. 
In conclusion, the 
$^3$He yield is within the statistical uncertainties of the alpha-particle yield 
for alpha energies 
between 4 and 8 MeV. A consequence of this procedure is that the $^3$He spectra 
have a low-energy cutoff 
of about 8 MeV.

\subsection{Low-energy neutron rejection and background subtraction}
\label{subsec:Low-energy}

Knowing the energy calibration and the flight distances, the flight time for 
each charged particle from 
target to detector can be calculated and subtracted from the registered total 
TOF. The resulting 
neutron TOF is used for selection of charged-particle events induced by neutrons 
in the main peak 
of the incident neutron spectrum. The TOF cut reduces the background of charged 
particles produced 
by peak neutrons hitting the chamber and telescope housing since the flight 
paths are different, 
especially for the backward telescopes. The widths of the TOF cuts in all 
detectors are fixed to 3$\sigma$ 
where $\sigma$ is the standard deviation of the H(n,p) peak in the $20^\circ$ 
telescope. Fig.~\ref{fig:fig2}a 
illustrates the selection procedure for deuterons at $20^\circ$ laboratory 
angle. The solid line is a 
kinematic calculation of the ground state peak in the deuteron spectra for each 
corresponding neutron 
energy. It provides a cross check of the energy and time calibration of the 
whole energy spectrum. 

Background events, measured in target-out runs and analyzed in the same way as 
target-in events, are 
subtracted from the corresponding target-in runs after normalization to the same 
neutron fluence. 
Fig.~\ref{fig:fig2}b shows the resulting spectrum of deuteron events at 
$20^\circ$ induced by the main 
neutron peak. For comparison, the same spectrum without TOF cut is presented. 
Finally, the target-out 
background, obtained with the same TOF cut is shown. The signal-to-background 
ratio is about 4. 

\subsection{Absolute cross section normalization}
\label{subsec:Absolute cross}

Absolute double-differential cross sections are obtained by normalizing the 
silicon data to the 
number of recoil protons emerging from the CH$_2$ target. After selection of 
events in the main 
neutron peak and proper subtraction of the target-out and $^{12}$C(n,p) 
background contributions, the 
latter taken from a previous experiment, the cross section can be determined 
from the recoil proton 
peak, using $np$ scattering data~\cite{Rah01}. All data have been normalized 
using the $np$ scattering 
peak in the $20^\circ$ telescope. As a cross check, Si(n,px) spectra have also 
been normalized using 
the $np$ scattering peak in the $40^\circ$ and $60^\circ$ telescopes, resulting 
in 
spectra in agreement with those normalized to the $20^\circ$ telescope.


\section{Corrections}
\label{sec:Corrections}


\subsection{Thick target correction}
\label{subsec:Thick target}

Due to the thickness of the target and to the low-energy cutoffs in the particle 
identification, the 
measured low-energy charged particles are produced in fractions of the entire 
thickness of the target. 
Therefore, not only energy-loss corrections are needed but also particle-loss 
corrections. Charged 
particles with the initial kinetic energy ${E_{init}}$ have a well-defined range 
$R$ in the target material. 
If ${R(E_{init})}$ is equal to or larger than the target thickness, all produced 
particles can escape from 
the target and no particle loss correction is required. If, on the other hand, 
${R(E_{init})}$ is smaller 
than the target thickness, a correction for particles stopped inside the target 
is needed.

The adopted correction method employs an initial energy ${(E_{init})}$ 
distribution called the inverse 
response function for each measured energy. For the 303 $\mu$m silicon target 
used in the present 
experiment, a measured alpha particle of 4 MeV could either be due to a 4 MeV 
particle from the front 
surface of the target, a 27 MeV particle from the back surface of the target, or 
anything in between. 
Therefore, the content of the measured energy bin should be redistributed over 
the initial energy 
region from 4 to 27 MeV. A FORTRAN program, TCORR~\cite{Pomp}, has been 
developed which calculates the 
inverse response functions, initially assuming an energy-independent cross 
section. These inverse response 
functions are normalized to the corresponding bin content in the measured 
spectrum and summed to get 
the true initial energy spectrum. Finally, the particle loss correction is 
applied. The resulting 
spectrum is folded with the primary inverse response functions to get improved 
inverse response functions 
and the procedure is repeated. Resulting spectra from two successive iterations 
are compared by a 
Kolmogorov test~\cite{HBOOK} to judge the convergence.

Results from the correction method have been verified with an independent 
Monte-Carlo program called 
TARGSIM, based on the GEANT code~\cite{GEANT}. This program simulates the 
measured spectra using the 
corrected spectra and the MEDLEY geometry as input. The simulation results are 
in agreement with the 
experimental data within the statistical errors over the whole energy region.

Obviously, the simulated spectra have much better statistics than the original 
experimental spectra and, 
therefore, the statistical fluctuations between neighboring energy bins are much 
smaller. In a sense, 
they are fits to the experimental spectra. In order to estimate the systematic 
uncertainty introduced 
by the thick target correction, these simulated spectra are corrected with TCORR 
again and compared with 
the result of the first correction. The observed differences for individual 
energy bins are typically 
5 \% and less than 10 \% in general. An extreme value of 40 \% was found in the 
lowest bin of the 
alpha spectrum at $140^\circ$. However, in all cases, except for protons where 
the statistical errors 
are very small, the deviations are within the statistical uncertainties of the 
original corrected data.

In conclusion, the systematic error of the target correction comes essentially 
from the statistical 
uncertainties. For protons and deuterons we estimate that this error is about 10 
\% in the lowest two 
energy bins decreasing to a few percent from 15 MeV and upwards. Due to less 
statistics and the 
increasing width of the inverse response functions, the uncertainty is larger 
for tritons, $^3$He and 
alpha particles, where it is 20 \% in the lowest two bins, decreasing to 10 \% 
above 25 MeV.

In addition, evaluated data~\cite{Watanabe} were used as input to check the 
reliability of our 
programs, obviously because validation with known "realistic" data is desirable. 
The latter have been 
simulated with the TARGSIM program to get pseudo-experimental data and have 
subsequently been corrected 
with the TCORR program using the same conditions as in the experiment. The 
corrected results appear to 
reproduce the known "realistic" data well.

\subsection{Collimator correction}
\label{subsec:Collimator correction}

As mentioned in sect.~\ref{sec:Experimental setup}, active collimators have been 
placed in front of the 
telescope in order to define the solid angle. However, due to malfunctioning in 
the present experiment, 
the signal from these collimators could not be used to suppress events hitting 
them. Therefore, although 
the collimators actually work as passive collimators for helium particles below 
35 MeV, their effect when 
particles punch through them has to be corrected for. To this end, a FORTRAN 
program has been developed 
that, based on the measured spectrum of particles and an iteration procedure, 
estimates the shape and 
fraction of the energy spectrum of particles hitting the collimator. It has been 
found that the corrections 
in shape are rather small and under control in all cases. The systematic error 
related to this correction 
comes from the uncertainty in the solid angle subtended by the silicon detectors 
for high-energy protons 
relative to the solid angle subtended by the collimator opening. This 
uncertainty is estimated to be 5 \% 
and, due to the normalization procedure, only affects the helium spectra and the 
low-energy part of the 
hydrogen spectra.

\subsection{Other corrections}
\label{subsec:Other corrections}

The 17 MHz repetition rate of the cyclotron beam pulse, which limits the TOF 
window to 58 ns, causes 
wrap-around problems. Thus, it is not possible to distinguish 96 MeV neutrons 
from those of 26 MeV 
created by the previous beam burst, since the latter have the same apparent TOF. 
 This can be seen in 
Fig.~\ref{fig:fig2}a, where the bent band from the low-energy neutron tail 
crosses the straight band of the 
full-energy neutrons. Since the Q-value for the $^{28}$Si(n,d) reaction is 
$-9.4$ MeV, this interference shows 
up as a bump below 20 MeV in Fig.~\ref{fig:fig2}b. A correction for this 
effected is applied, using tabulated
values from Ref.~\cite{ICRU} and a  ratio of the neutron fluence in the 
wrap-around region and at 96 MeV of 6.3 \%.
For the neutron-energy spectrum in the wrap-around region, a square distribution 
ranging from 24 to 29 MeV is assumed.
Thus, the cross sections at 
24, 26 and 28 MeV as given in Ref.~\cite{ICRU} are used, with the 24 MeV values 
entering at half weight.
The data for 80, 100, 120, 140 and 160 degrees are obtained by linear 
interpolation. Only the spectra 
for proton, deuteron and alpha particles are corrected. The effect of this 
correction is a reduction of about 5 \% in 
the production cross section. 
In the (n,d) spectra presented in Fig.~\ref{fig:fig4}, some structure at
$20^\circ$ and $40^\circ$ around 15 MeV might be attributed to deficiencies in the correction.
The triton production cross sections given in 
Ref.~\cite{ICRU}
(Q-value = -16.2 MeV) indicate, that the correction would be about an order of 
magnitude lower and is therefore negligible. 
For $^3$He production (Q-value = -12.1 MeV), the correction 
is also negligible due to 
the high-energy cutoff of 8 MeV.

There is a TOF shift problem, seen as a band parallel with the main band in 
Fig.~\ref{fig:fig2}a. The reason 
for this is probably that the electronic timing module has not worked properly. 
This is corrected by extending 
the TOF cut with the dotted rectangle in the same figure, to include these 
events. This method could be 
applied only in the energy region where there is no interference from the low 
energy neutron tail. Therefore, 
the ratio of the number of events between the parallel and the main band is 
determined and then applied to 
the low-energy region as well. This ratio is 1.3  \% in the worst case.

Albeit a majority of the neutrons appears in the narrow full-energy peak at 95.6 
MeV, a significant fraction 
(about 25  \%) belongs to a tail extending towards lower energies, remaining 
also after the TOF cut. The 
average neutron energy with these tail neutrons included is 92.4 MeV. This 
effect has been taken into account 
in the normalization of the data.

Minor corrections of a few percent are applied to the experimental spectra for 
the CsI(Tl) intrinsic 
efficiency~\cite{Klu01} and for the dead time in the data acquisition system.


\section{Theoretical models}
\label{sec:Theoretical models}


Data have been compared with nuclear theory predictions, computed with the two 
nuclear reaction codes 
GNASH~\cite{Young,Cha97} and TALYS~\cite{Koning}. While GNASH has been widely 
used during the last years, 
TALYS is a new code still under development. Two sets of GNASH calculations are 
presented, one with 
parameters as presented in a recent evaluation~\cite{ICRU}, and another set with 
modified parameters 
~\cite{Watanabe} as described in sect.~\ref{subsec:GNASH calculations}. The 
latter parameter set is 
developed as part of another data evaluation~\cite{Sun02}. Since the latter work 
and TALYS are not 
published, they are described in some detail below.

Both GNASH and TALYS integrate direct, pre-equilibrium, and statistical nuclear 
reaction models into one 
calculation scheme and thereby give predictions for all the open reaction 
channels. Both codes use the 
Hauser-Feshbach model for sequential equilibrium decay and the exciton model for 
pre-equilibrium 
emission. The angular distributions are obtained using the Kalbach 
systematics~\cite{Kal88}.

\subsection{TALYS calculations}
\label{subsec:TALYS calculations}

The purpose of TALYS is to simulate nuclear reactions that involve neutrons, 
photons, protons, deuterons, 
tritons, $^3$He and alpha particles, in the 1 keV -- 200 MeV energy range. 
Predicted quantities include 
integrated, single- and double-differential cross sections, for both the 
continuum and discrete states, 
residue production and fission cross sections, gamma-ray production cross 
sections, etc. For the present 
work, single- and double-differential cross sections are of interest. To predict 
these, a calculation 
scheme is invoked which consists of a direct + pre-equilibrium reaction 
calculation followed by subsequent 
compound nucleus decay of all possible residual nuclides calculated by means of 
the Hauser-Feshbach model.

First, dedicated optical model potentials (OMP) were developed for both neutrons 
and protons on $^{28}$Si 
up to 200 MeV. The used parameters are from the OMP collection of 
Ref.~\cite{Kon03}. These potentials 
provide the necessary reaction cross sections and transmission coefficients for 
the statistical model 
calculations. For complex particles, the optical potentials were directly 
derived from the nucleon 
potentials using the folding approach of Watanabe~\cite{Wat58}. 

Pre-equilibrium emission takes place after the first stage of the reaction but 
long before statistical 
equilibrium of the compound nucleus is attained. It is imagined that the 
incident particle step-by-step 
creates more complex states in the compound system and gradually loses its 
memory of the initial energy 
and direction. The default pre-equilibrium model of TALYS is the two-component 
exciton model of Kalbach 
~\cite{Kal86}. In the exciton model (see Refs.~\cite{Gru86} and~\cite{Gad92} for 
extensive reviews), at 
any moment during the reaction, the nuclear state is characterized by the total 
energy $E_{tot}$ and the 
total number of particles above and holes below the Fermi surface. Particles 
($p$) and holes ($h$) are 
indiscriminately referred to as excitons. Furthermore, it is assumed that all 
possible ways of sharing 
the excitation energy between different particle-hole configurations with the 
same exciton number $n=p+h$ 
have equal probability. To keep track of the evolution of the scattering 
process, one merely traces the 
development of the exciton number, which changes in time as a result of 
intranuclear two-body collisions. 

The basic starting point of the exciton model is a time-dependent master 
equation, which describes the 
probability of transitions to more and less complex particle-hole states as well 
as transitions to the 
continuum, i.e., emission. Upon integration over time, the energy-averaged 
emission spectrum is obtained. 
The assumptions above make the exciton model amenable to practical calculations. 
This, however, requires 
the introduction of a free parameter, namely the average matrix element of the 
residual two-body 
interaction, occurring in the transition rates between two exciton states. 
Without going into details, 
the basic formulae are given for the two-component exciton model. The created 
particles and holes of 
proton and neutron type are explicitly followed throughout the reaction. A 
notation is used in which 
$p_{\pi} (p_{\nu})$ is the proton (neutron) particle number and $h_{\pi} 
(h_{\nu})$ the proton (neutron) 
hole number. Following Kalbach~\cite{Kal86}, the exciton model cross section is 
now given by 
\begin{equation}
\frac{d\sigma_{k}^{EM}}{dE_{k}} = \sigma^{CF}
\sum_{p_{\pi}=p_{\pi}^{0}}^{p_{\pi}^{eq}}
\sum_{p_{\nu}=p_{\nu}^{0}}^{p_{\nu}^{eq}}
w_{k}(p_{\pi},h_{\pi},p_{\nu},h_{\nu},E_{k})
S_{pre}(p_{\pi},h_{\pi},p_{\nu},h_{\nu}),
\label{eq:pespec2}
\end{equation}
where $\sigma^{CF}$ is the compound formation cross section and $S_{pre}$ the 
time-integrated strength which 
determines how long the system remains in a certain exciton 
configuration~\cite{Kal86}. The initial proton 
and neutron particle numbers are denoted $p_{\pi}^{0}=Z_{p}$,and 
$p_{\nu}^{0}=N_{p}$ with $Z_{p} (N_{p})$ 
being the proton (neutron) number of the projectile. In general, 
$h_{\pi}=p_{\pi}-p_{\pi}^{0}$ and $h_{\nu}=
p_{\nu}-p_{\nu}^{0}$, so that the initial hole numbers are zero, i.e. 
$h_{\pi}^{0}=h_{\nu}^{0}=0$, for primary 
pre-equilibrium emission. The pre-equilibrium part is calculated by 
Eq.~\ref{eq:pespec2}, using $p_{\pi}^{eq}=p_{\nu}^{eq}=6$, 
whereas the remainder of the reaction flux is distributed through the 
Hauser-Feshbach model. In addition, the 
never-come-back approximation is adopted.

The emission rate $w_k$ for ejectile $k$ with spin $s_k$ is given by
\begin{equation}
w_{k}(p_{\pi},h_{\pi},p_{\nu},h_{\nu},E_{k}) = \frac{2s_{k}+1}{\pi ^{2}\hbar ^{3
}}\mu _{k} E_{k}
\sigma _{k,inv}(E_{k})
\frac{\omega (p_{\pi}-Z_{k},h_{\pi},p_{\nu}-N_{k},h_{\nu},E_{x})}
{\omega (p_{\pi},h_{\pi},p_{\nu},h_{\nu},E^{tot})},
\label{eq:emission2}
\end{equation}
where $\sigma _{k,inv}(E_{k})$ is the inverse reaction cross section as 
calculated from the optical model 
and $\omega$ is the two-component particle-hole state density. The expression 
for $S_{pre}$ contains the 
adjustable transition matrix element $M^{2}$ for each possible transition 
between neutron-proton exciton 
configurations. A proton-neutron ratio of 1.6 for the squared internal 
transition matrix elements was adopted 
to give the best overall agreement with experiment, i.e., 
$M_{\pi\nu}^{2}=M_{\nu\pi}^{2}=1.6M_{\pi\pi}^{2}=
1.6M_{\nu\nu}^{2}$. Partial level density parameters $g_{\pi}=Z/15$ and 
$g_{\nu}=N/15$ were used in the 
equidistant spacing model for the partial level densities.

At incident energies above several tens of MeV, the residual nuclides formed 
after binary emission may 
have so large excitation energy that the presence of additional fast particles 
inside the nucleus becomes 
possible. The latter can be imagined as strongly excited particle-hole pairs 
resulting from the first 
binary interaction with the projectile. The residual system is then clearly 
non-equilibrated and the 
excited particle that is high in the continuum may, in addition to the first 
emitted particle, also be 
emitted on a short time scale. This so-called multiple pre-equilibrium emission 
forms an alternative 
theoretical picture of the intra-nuclear cascade process, whereby the exact 
location and momentum of 
the particles are not followed, but instead the total energy of the system and 
the number of 
particle-hole excitations (exciton number).

In actual calculations, the particle-hole configuration of the residual nucleus 
after emission of the 
ejectile, is re-entered as initial condition in Eq.~\ref{eq:pespec2}. When 
looping over all possible residual 
configurations, the multiple pre-equilibrium contribution is obtained. In TALYS, 
multiple pre-equilibrium 
emission is followed up to arbitrary order, though for 96 MeV only secondary 
pre-equilibrium emission is 
significant. 

It is well-known that semi-classical models, such as the exciton model, have 
always had some problems to 
describe angular distributions (essentially because it is based on a 
compound-like concept instead of a 
direct one). Therefore, as mentioned previously, the double-differential cross 
sections are obtained from 
the calculated energy spectra using the Kalbach systematics~\cite{Kal88}. 

To account for the evaporation peaks in the charged-particle spectra, multiple 
compound emission was 
treated with the Hauser-Feshbach model. In this scheme, all reaction chains are 
followed until all 
emission channels are closed. The Ignatyuk model~\cite{Ign75} has been adopted 
for the total level 
density to account for the damping of shell effects at high excitation energies.

For pre-equilibrium reactions involving deuterons, tritons, $^3$He and alpha 
particles, a contribution 
from the exciton model is automatically calculated with the formalism described 
above. It is, however, 
well known that for nuclear reactions involving projectiles and ejectiles with 
different particle numbers, 
mechanisms like stripping, pick-up and knock-out play an important role and 
these direct-like reactions 
are not covered by the exciton model. Therefore, Kalbach developed a 
phenomenological contribution for 
these mechanisms~\cite{Kal01}, which is included in TALYS. It has recently been 
shown (see Table I of Ref.~\cite{Ker02}) 
that this method gives a considerable improvement over the older methods. The 
latter seemed to consistently 
underpredict neutron-induced reaction cross sections.

\subsection{GNASH calculations}
\label{subsec:GNASH calculations}

For the present work, GNASH calculations have been performed with a modified 
parameter set. The 
calculation procedure is outlined in Ref.~\cite{Sun02}. Transmission 
coefficients needed for the GNASH 
input were calculated using the optical potential parameters by Sun {\it et 
al.}~\cite{Sun03} for neutrons 
and protons, Daehnick {\it et al.}~\cite{Dae80} for deuterons, 
Becchetti-Greenlees~\cite{Bec71} for tritons 
and $^3$He particles, and Avrigeanu {\it et al.}~\cite{Avr94} for alpha 
particles.

Like in the TALYS case, default level density parameters were used with the 
Ignatyuk level density 
formula~\cite{Ign75}. The normalization factor used in the pre-equilibrium model 
calculation was 
determined by analyses of proton-induced reactions. The calculated result of 
pre-equilibrium deuteron and 
alpha emission is different from that of the original GNASH code 
calculation~\cite{ICRU}. In the 
deuteron emission, the component with the exciton number 3 was ignored. The 
direct pick-up component was 
calculated using a phenomenological approach~\cite{Kal77} with a normalization 
that is independent of 
the incident energy. This normalization was determined from analysis of 
experimental (n,dx) energy 
spectra up to 60 MeV~\cite{Bat99}. The alpha knockout component given by the 
same phenomenology~\cite{Kal77} was ignored. 

The results are given in the laboratory system. Like in the TALYS case, angular 
distributions are 
obtained using the Kalbach systematics~\cite{Kal88}. The required 
pre-equilibrium fraction is taken from 
the GNASH output. The c.m.-to-lab transformation is performed using the 
kinematics of one-particle 
emission as described in Refs.~\cite{Young,Cha97}.

The exciton model implemented in GNASH is a one-component exciton model 
developed by Kalbach~\cite{Kal85}, 
with a parameterisation for the energy dependence of the squared internal 
transition matrix element which 
has been validated at relatively low incident energies (below 40 MeV). There are 
indications that at 
higher incident energies, this energy dependence is no longer appropriate and 
that a more general form, 
covering a wider energy range, is needed. Such a smooth form has been 
implemented in TALYS, on the basis 
of a collection of double-differential (nucleon-in,nucleon-out) cross section 
measurements~\cite{Koning}.


\section{Results and discussion}
\label{sec:Results and discussion}


Double-differential cross sections at laboratory angles of $20^\circ$, 
$40^\circ$, $100^\circ$ and 
$140^\circ$ for protons, deuterons, tritons, $^3$He and alpha particles are 
shown in Figs.~\ref{fig:fig3} -~\ref{fig:fig7}, 
respectively. All spectra for each particle type are plotted on the same cross 
section scale to 
facilitate the comparison of their magnitude. The choice of the energy bin width 
is a 
compromise between the energy resolution in the experiment, the width of the 
inverse response 
functions and acceptable statistics in each energy bin. The error bars represent 
statistical 
uncertainties only.

The overall relative statistical uncertainties of individual points in the 
double-differential energy 
spectra at $20^\circ$ are typically 3 \% for protons, 7 \% for deuterons, 20 \% 
for tritons, 20 \% 
for $^3$He and 15 \% for alpha particles. As the angular distributions are 
forward-peaked, these values 
increase with angle. The systematic uncertainty contributions are due to thick 
target correction 
($1-20$ \%), collimated solid angle ($1-5$ \%), beam monitoring ($2-3$ \%), the 
number of silicon nuclei 
(1 \%), CsI(Tl) intrinsic efficiency (1 \%), particle identification (1 \%) and 
dead time ($<$\,0.1 \%). 
The uncertainty in the absolute cross section is about 5 \%, which is due to 
uncertainties in the $np$ 
scattering angle, the contribution from the low-energy continuum of the 
$^7$Li(p,n) spectrum to the $np$ 
scattering proton peak (3 \%), the reference $np$ cross sections (2 
\%)~\cite{Rah01}, statistics in the 
$np$ scattering proton peak (2 \%), the carbon contribution (0.1 \%) and the 
number of hydrogen 
nuclei (0.1 \%).

From Figs.~\ref{fig:fig3} -~\ref{fig:fig7} it is obvious that the 
charged-particle emission from 96 MeV 
neutron irradiation of silicon is dominated by proton, deuteron and alpha 
particle channels. The spectra 
of the two other particle types studied in this work (tritons and $^3$He) are 
more than an order of 
magnitude weaker. All the spectra have more or less pronounced peaks at low 
energies (below $10-15$ MeV), 
the angular distributions of which are not too far from isotropy. This 
low-energy peak is not observed in 
the $^3$He spectra due to the 8 MeV low-energy cutoff discussed in 
sect.~\ref{subsec:Particle identification}.

All the particle spectra at forward angles show relatively large yields at 
medium-to-high energies. 
The emission of high-energy particles is strongly forward-peaked and hardly 
visible in the backward 
hemisphere. It is a sign of particle emission before statistical equilibrium has 
been reached in the 
reaction process. In addition to this broad distribution of emitted particles, 
the deuteron spectra at 
forward angles show narrow peaks corresponding to transitions to the ground 
state and low-lying states 
in the final nucleus, $^{27}$Al. These transitions are most likely due to 
pick-up of weakly bound protons 
in the target nucleus, $^{28}$Si. 

\subsection{Comparison with theoretical model calculations}
\label{subsec:Comparison with}

In Figs.~\ref{fig:fig3} -~\ref{fig:fig8} the experimental results are presented 
together with theoretical 
model calculations.
The GNASH calculations of Ref.~\cite{ICRU} have been done for protons, deuterons 
and alpha particles,
whereas the other two calculations have been performed for all five particle 
types.

Fig.~\ref{fig:fig3} shows a comparison between the double-differential (n,px) 
experimental spectra and the 
calculations based on the TALYS and GNASH models. For protons above 25 MeV, all 
calculations give a good 
description of the spectra. Below this energy, some differences can be observed, 
e.g., at forward angles 
TALYS gives a better description of the statistical peak than the GNASH 
calculations.

The situation is quite different for the deuteron spectra (Fig.~\ref{fig:fig4}). 
None of the predictions do 
account for the data. At all angles deviations of a factor of two or more are 
present. At forward angles 
the high-energy part is strongly overestimated, indicating problems in the 
hole-strength treatment. There 
is a large difference in the spectral shapes calculated with the two versions of 
GNASH~\cite{Watanabe,ICRU}. 
This difference is due to the fact that emission from the configurations with 
exciton number 3 is 
neglected in the present GNASH calculations. This component is taken into 
account as a direct pickup 
component calculated with an empirical formula due to Kalbach~\cite{Kal77}.

For tritons (Fig.~\ref{fig:fig5}), the TALYS calculations give a slightly better 
description of the experimental 
data, whereas for $^3$He (Fig.~\ref{fig:fig6}) some large deviations can be 
observed. The TALYS calculations 
seem to account better for the spectrum shapes.

The overall description of the alpha particle spectra (Fig.~\ref{fig:fig7}) is 
fair. The GNASH calculations 
overpredict the high-energy data at forward angles, whereas the TALYS 
predictions are too large at backward 
angles.

The ability of the models to account for the low-energy peak caused by 
evaporation processes is not 
impressive. In general, the models tend to overpredict the cross sections. It 
should, however, be kept in 
mind that the peak maximum is close to (for $^3$He below) the low-energy cutoff, 
which complicates the 
comparison. Another complication in this context is that the c.m.-to-lab 
transformation of the calculated 
TALYS spectra could, at least in some cases, make a considerable difference. The 
GNASH cross sections are 
given in the lab system, but the c.m.-to-lab transformation is performed using 
the kinematics of 
one-particle emission~\cite{Young,Cha97}, which obviously is an approximation.

\subsection{Integrated spectra}
\label{subsec:Integrated spectra}

For each energy bin of the light-ion spectra, the experimental angular 
distribution is fitted by a simple 
two-parameter functional form, $a\exp(b\cos\theta)$~\cite{Kal88}.

This allows extrapolation of double-differential cross sections to very forward 
and very backward angles. 
In this way coverage of the full angular range is obtained. By integration of 
the angular distribution, 
energy-differential cross sections ($d\sigma /dE$) are obtained for each 
ejectile. These are shown in 
Fig.~\ref{fig:fig8} together with the theoretical calculations. All calculations 
are in good agreement with 
the proton experimental data over the whole energy range. In the cases of 
deuterons and alpha particles, the 
models overpredict the high-energy parts of the spectra.

The production cross sections are deduced by integration of the 
energy-differential spectra (see Table~\ref{tab:tab1}). 
As explained above, the experimental values in Table~\ref{tab:tab1} have to be 
corrected for the undetected 
particles below the low-energy cutoff. This is particularly important for $^3$He 
because of the high cutoff.

The proton and deuteron production cross sections are compared with previous 
data at lower energies~\cite{Ben02b} 
in Figs.~\ref{fig:fig9} and~\ref{fig:fig10}. There seems to be general agreement 
between the trend of the 
previous data and the present data point. The curves in these figures are based 
on a GNASH calculation~\cite{ICRU}.   


\section{Conclusions and outlook}
\label{sec:Conclusions and outlook}


In the present paper, we report an experimental data set on light-ion production 
induced by 96 MeV 
neutrons on silicon. Experimental double-differential cross sections 
($d^2\sigma/d\Omega dE$) are 
measured at eight angles between $20^\circ$ and $160^\circ$. Energy-differential 
($d\sigma /dE$) and 
production cross sections are obtained for the five types of outgoing particles. 
Theoretical calculations 
based on nuclear reaction codes including direct, pre-equilibrium and 
statistical calculations give 
generally a good account of the magnitude of the experimental cross sections. 
For proton emission, the 
shape of the spectra for the double-differential and energy-differential cross 
sections are well described. 
The calculated and the experimental alpha-particle spectra are also in fair 
agreement with the exception 
of the high energy part, where the theory predicts higher yields than 
experimentally observed. For the 
other complex ejectiles (deuteron, triton and $^3$He) there are important 
differences between theory and 
experiment in what concerns the shape of the spectra.

For the further development of the field, data at even higher energies are 
requested. The results 
suggest that the MEDLEY facility, which was used for the present work, should be 
upgraded to work also 
at 180 MeV, i.e., the maximum energy of the TSL neutron beam facility. At 
present, a new neutron beam 
facility is under commissioning at TSL, covering the same energy range, but with 
a projected intensity 
increase of a factor five. This will facilitate measurements at higher energies 
than in the present work.

The setup described in this paper comprises an active target, the information of 
which was not used in 
the analysis here but can provide valuable information on the kinetic energy 
transferred to the residual 
nucleus. This information might be crucial for soft-error studies and therefore 
it is of interest to 
compare this measurement with theoretical calculations. Work along this line is 
in progress.

\section*{Acknowledgments}

This work was supported by the Swedish Natural Science Research Council, the 
Swedish Nuclear Fuel and 
Waste Management Company, the Swedish Nuclear Power Inspectorate, Ringhals AB, 
and the Swedish Defence 
Research Agency. The authors wish to thank the The Svedberg Laboratory for 
excellent support. One of the 
authors (U.T.) wishes to express his gratitude to the Thai Ministry of 
University Affairs and to the 
International Program in the Physical Sciences at Uppsala University. One of the 
authors (Y.W.) is 
grateful to the scientific exchange program between the Japan Society for the 
Promotion of Science and 
the Royal Swedish Academy of Sciences.


\pagebreak

\section*{Figure captions}

\begin{enumerate}
\item
\label{fig:fig1}
(a) Particle identification spectra at $20^\circ$ for the $\Delta E_1-\Delta 
E_2$ (a) and $\Delta E_2-E$ 
(b) detector combinations. The solid lines represent tabulated energy loss 
values in silicon~\cite{Ziegler}. 
The insert in (b) illustrates the separation of high-energy protons, deuterons 
and tritons discussed in 
sect.~\ref{subsec:Particle identification}. 


\item
\label{fig:fig2}
(a) Neutron TOF spectrum versus deuteron energy for the Si(n,dx) reaction at 
$20^\circ$ and the selection of 
deuterons associated with the full-energy neutron peak. The neutron-energy scale 
is given to the right. The 
solid line is a kinematic calculation of the ground-state deuteron energy as a 
function of the neutron 
energy. The lower rectangular cut is associated with neutrons in the full-energy 
peak, whereas the adjacent 
rectangular cut is used when correcting for the observed timing shift discussed 
in sect.~\ref{subsec:Other corrections}.
(b) Deuteron energy spectrum at $20^\circ$ with (solid histogram) and without 
(dashed histogram) the 
full-energy neutron cut. The cross-hatched histogram shows the target-out 
background. The bump below 20 MeV in 
the solid histogram is due to wrap-around effects discussed in 
sect.~\ref{subsec:Other corrections}.


\item
\label{fig:fig3}
Experimental double-differential cross sections (filled circles) of the Si(n,px) 
reaction at 96 MeV at four 
laboratory angles. The curves indicate theoretical calculations based on 
GNASH~\cite{ICRU} (dashed), TALYS-present 
work (dotted) and GNASH-present work (solid). The TALYS result is in the c.m. 
system and the GNASH results are in 
the lab system.


\item
\label{fig:fig4}
Experimental double-differential cross sections (filled circles) of the Si(n,dx) 
reaction at 96 MeV at four 
laboratory angles. The curves indicate theoretical calculations based on 
GNASH~\cite{ICRU} (dashed), TALYS-present 
work (dotted) and GNASH-present work (solid). The TALYS result is in the c.m. 
system and the GNASH results are in 
the lab system.


\item
\label{fig:fig5}
Experimental double-differential cross sections (filled circles) of the Si(n,tx) 
reaction at 96 MeV at four 
laboratory angles. The curves indicate theoretical calculations based on 
TALYS-present 
work (dotted) and GNASH-present work (solid). The TALYS result is in the c.m. 
system and the GNASH result is in 
the lab system.


\item
\label{fig:fig6}
Experimental double-differential cross sections (filled circles) of the 
Si(n,$^3$Hex) reaction at 96 MeV at four 
laboratory angles. The curves indicate theoretical calculations based on 
TALYS-present 
work (dotted) and GNASH-present work (solid). The TALYS result is in the c.m. 
system and the GNASH result is in 
the lab system.


\item
\label{fig:fig7}
Experimental double-differential cross sections (filled circles) of the 
Si(n,$\alpha$x) reaction at 96 MeV at four 
laboratory angles. The curves indicate theoretical calculations based on 
GNASH~\cite{ICRU} (dashed), TALYS-present 
work (dotted) and GNASH-present work (solid). The TALYS result is in the c.m. 
system and the GNASH results are in 
the lab system.
Note the logarithmic scale.


\item
\label{fig:fig8}
Experimental energy-differential cross sections (filled circles) for 
neutron-induced p, d, t, $^3$He and 
$\alpha$ production at 96 MeV. The curves indicate theoretical calculations 
based on GNASH~\cite{ICRU}
(dashed), TALYS-present work (dotted) and GNASH-present work (solid). The TALYS 
result is in the c.m. system 
and the GNASH results are in the lab system.


\item
\label{fig:fig9}
Neutron-induced proton production cross section as a function of neutron energy. 
The full circle is from 
the present work, whereas the open circles are from previous work~\cite{Ben02a}. 
The curve is based on a 
GNASH calculation~\cite{ICRU}. The data as well as the calculations correspond 
to a cutoff energy of 4 MeV.
Note that the cutoff energy is different from that in Table ~\ref{tab:tab1}.


\item
\label{fig:fig10}
Same as Fig.~\ref{fig:fig9} for deuteron production, with a cutoff energy of 8 
MeV.

\end{enumerate}

\pagebreak

\begin{table}
\caption{Experimental production cross sections for protons, deuterons, tritons, 
$^3$He 
and alpha particles from the present work. Theoretical values resulting from 
GNASH and TALYS calculations are 
given as well. The experimental data in the second column have been obtained 
with cutoff energies of 2.5, 3.0, 3.5, 
8.0 and 4.0 MeV for p, d, t, $^3$He and alpha particles, respectively. The third 
column shows data corrected for 
these cutoffs, using the GNASH calculation of the present work.
}
\label{tab:tab1} 
\begin{tabular}{lccccc} 
\hline
\hline
$\sigma_{prod}$ & Experiment & Experiment & GNASH & GNASH & TALYS \\
 & (mb)& (cutoff corr.) & (Ref.~\protect\cite{ICRU}) & (present) & (present)\\
\hline
(n,px) & 436$\pm$22 & 507 & 670.3 & 701.9 & 558.3\\ 
(n,dx) & 81$\pm$4 & 89.5 & 77.0 & 109.6 & 107.6\\ 
(n,tx) & 15.2$\pm$0.8 & 17.9 & -- & 15.0 & 13.1\\ 
(n,$^3$Hex) & 7.8$\pm$0.5 & 13.0 & -- & 10.6 & 14.5\\ 
(n,$\alpha$x) & 144$\pm$7 & 183 & 175.8 & 202.4 & 146.8\\ 
\hline
\end{tabular}
\end{table}


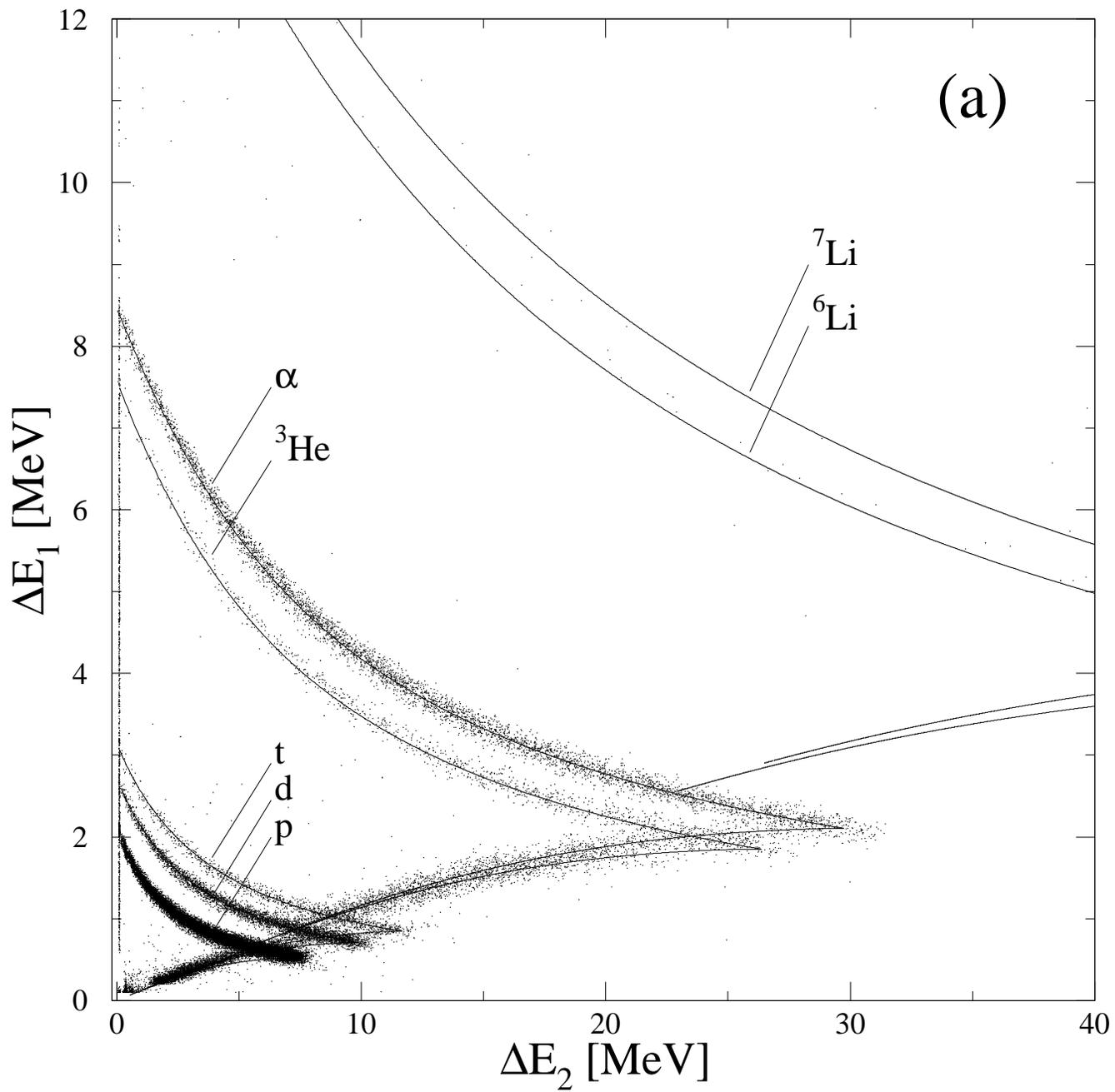
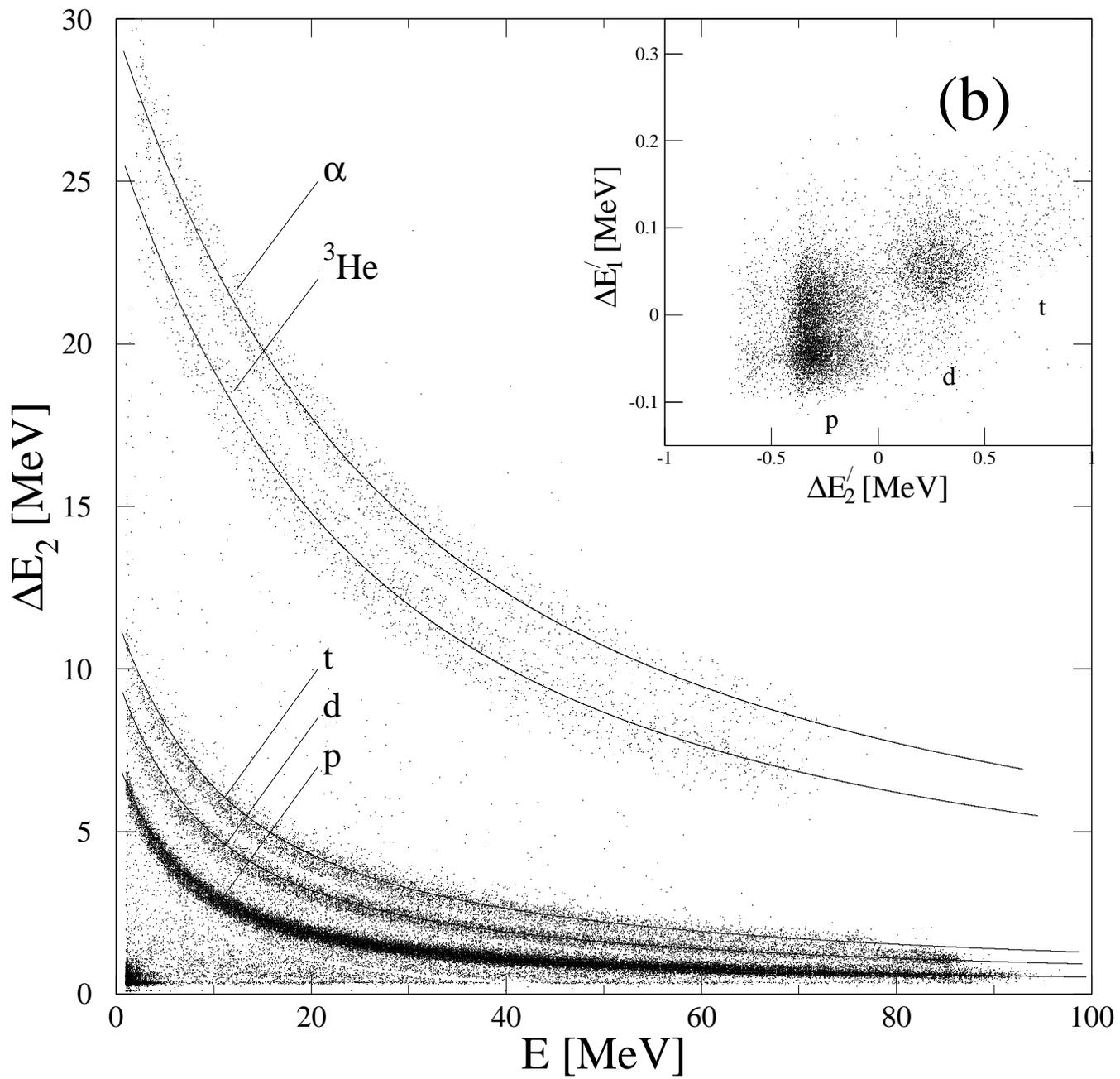
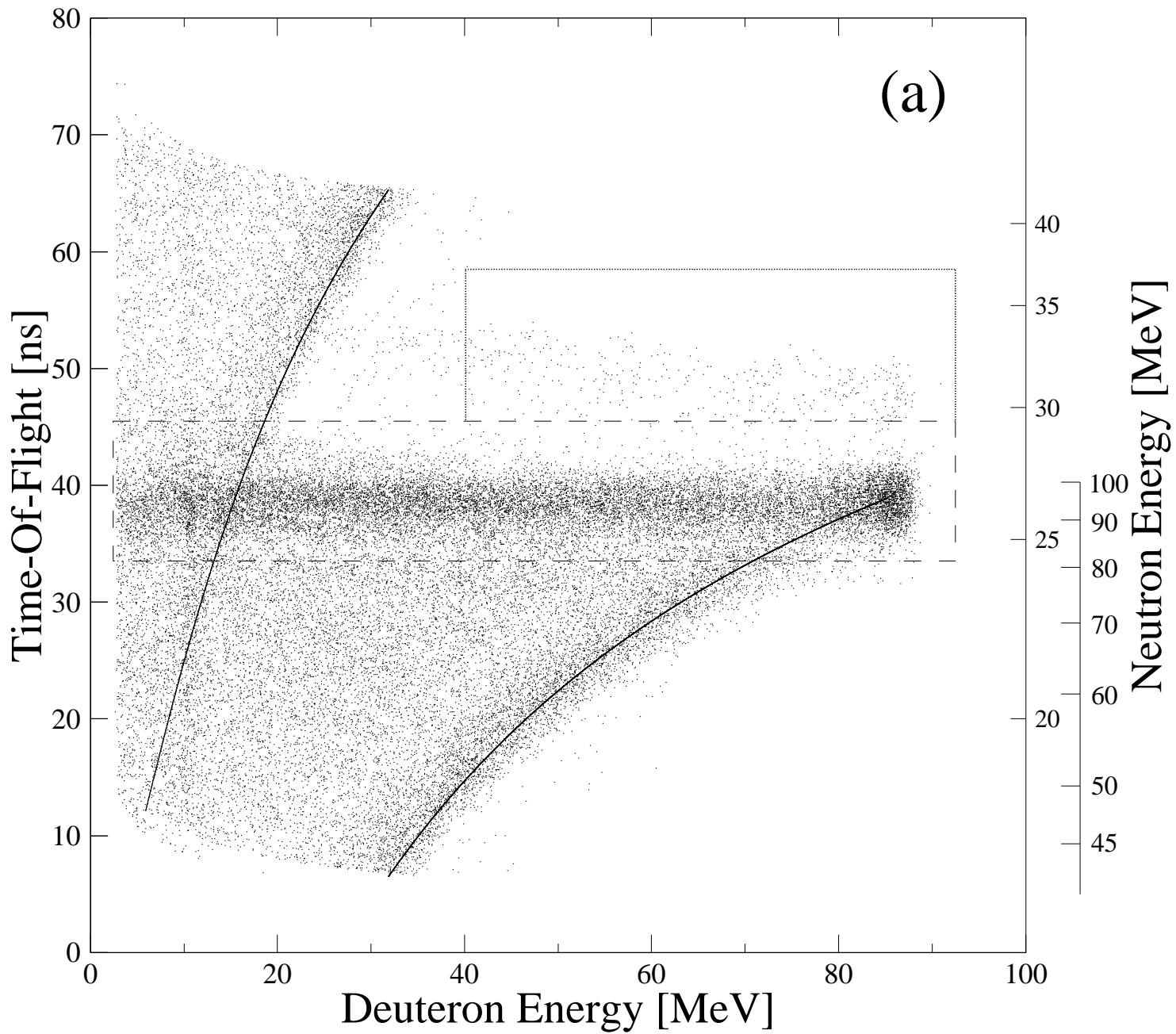
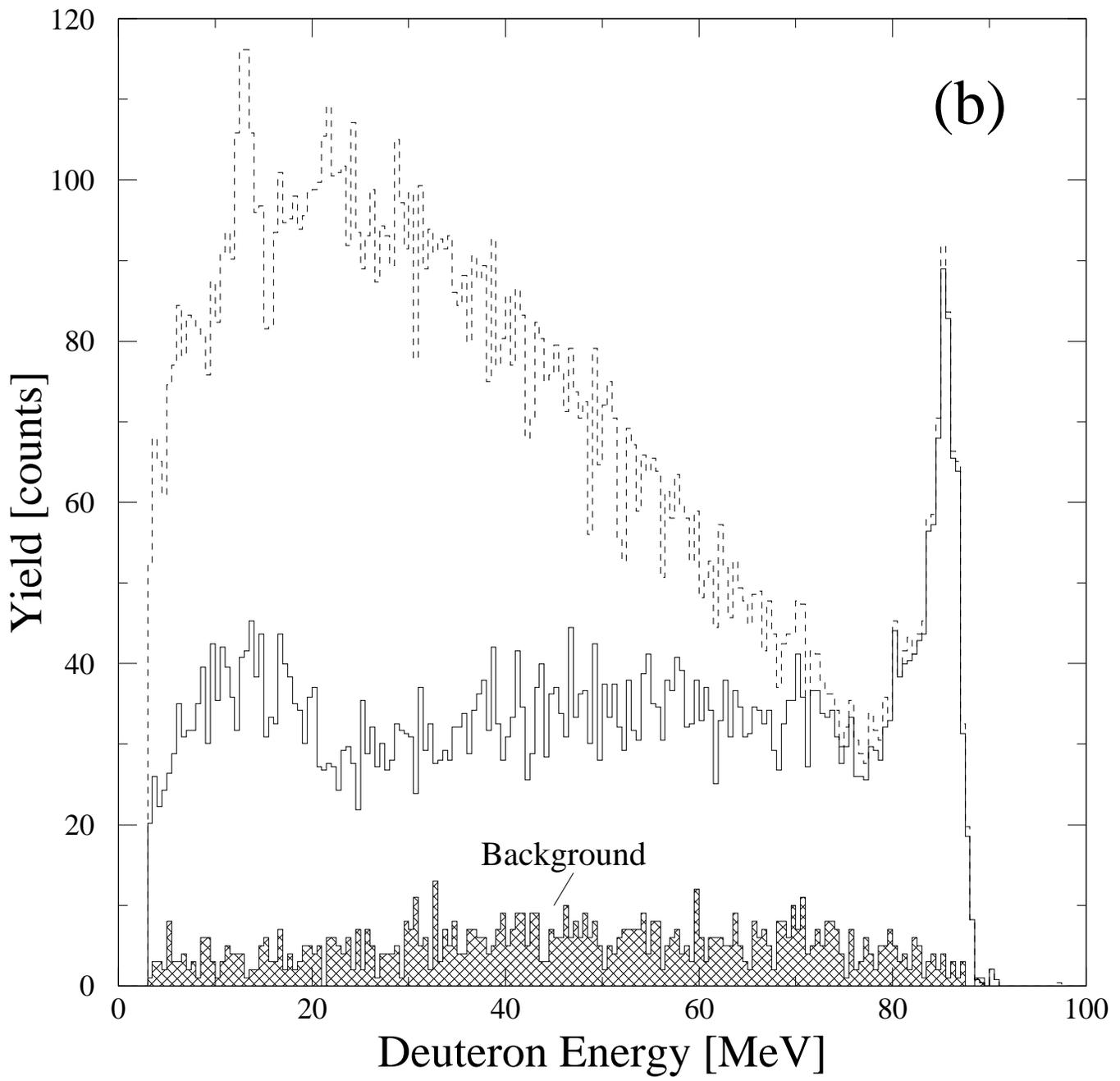
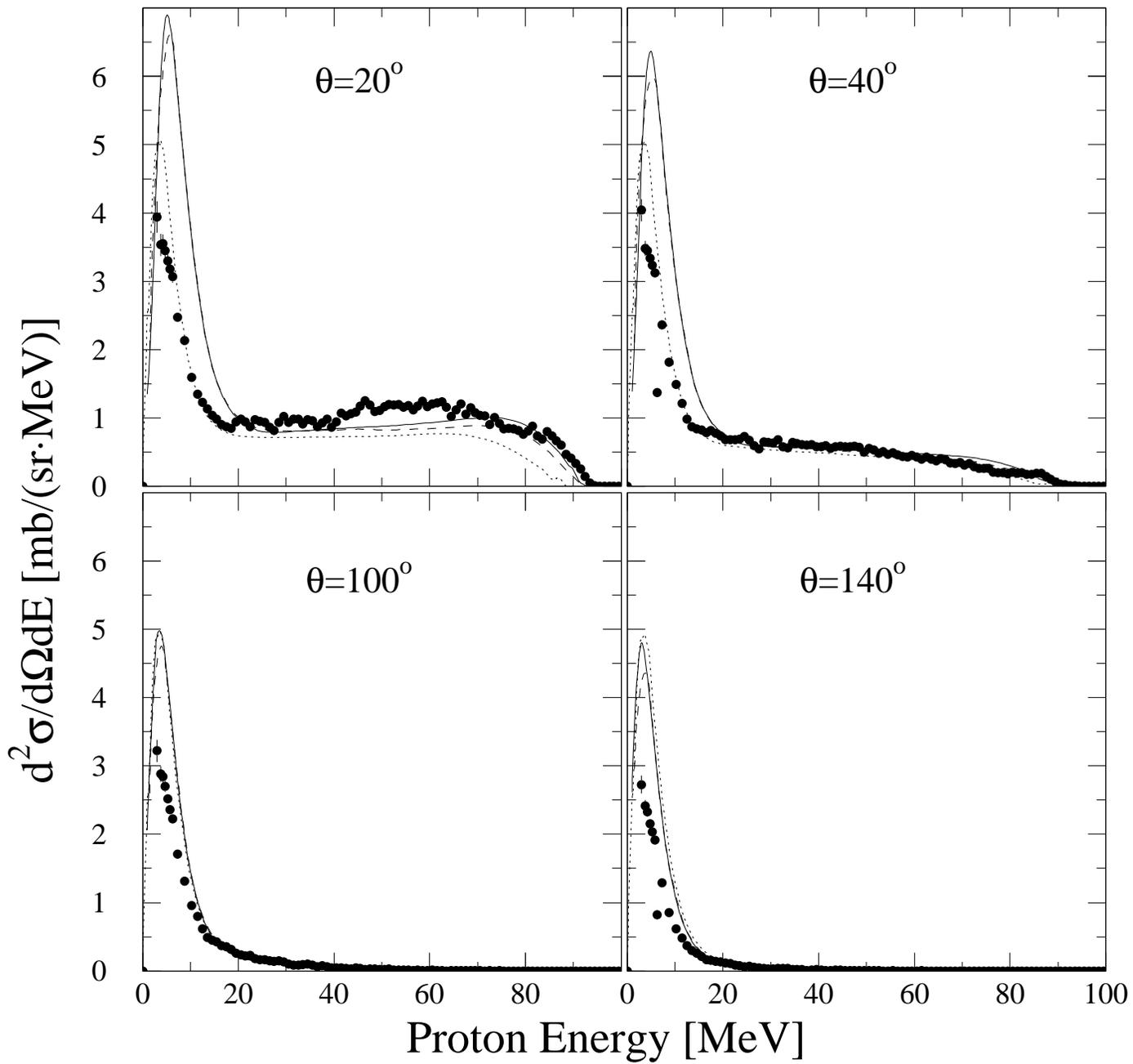
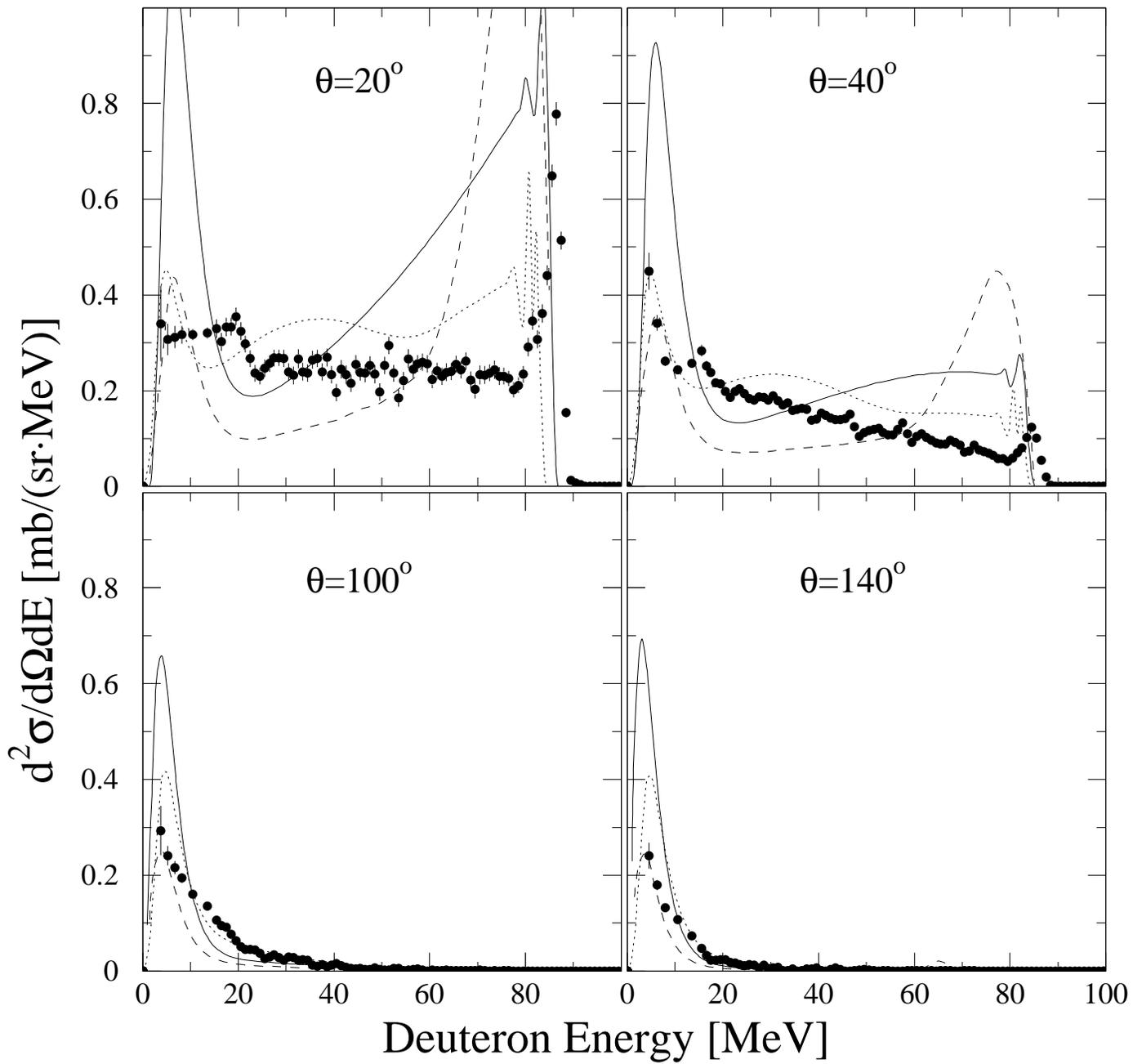
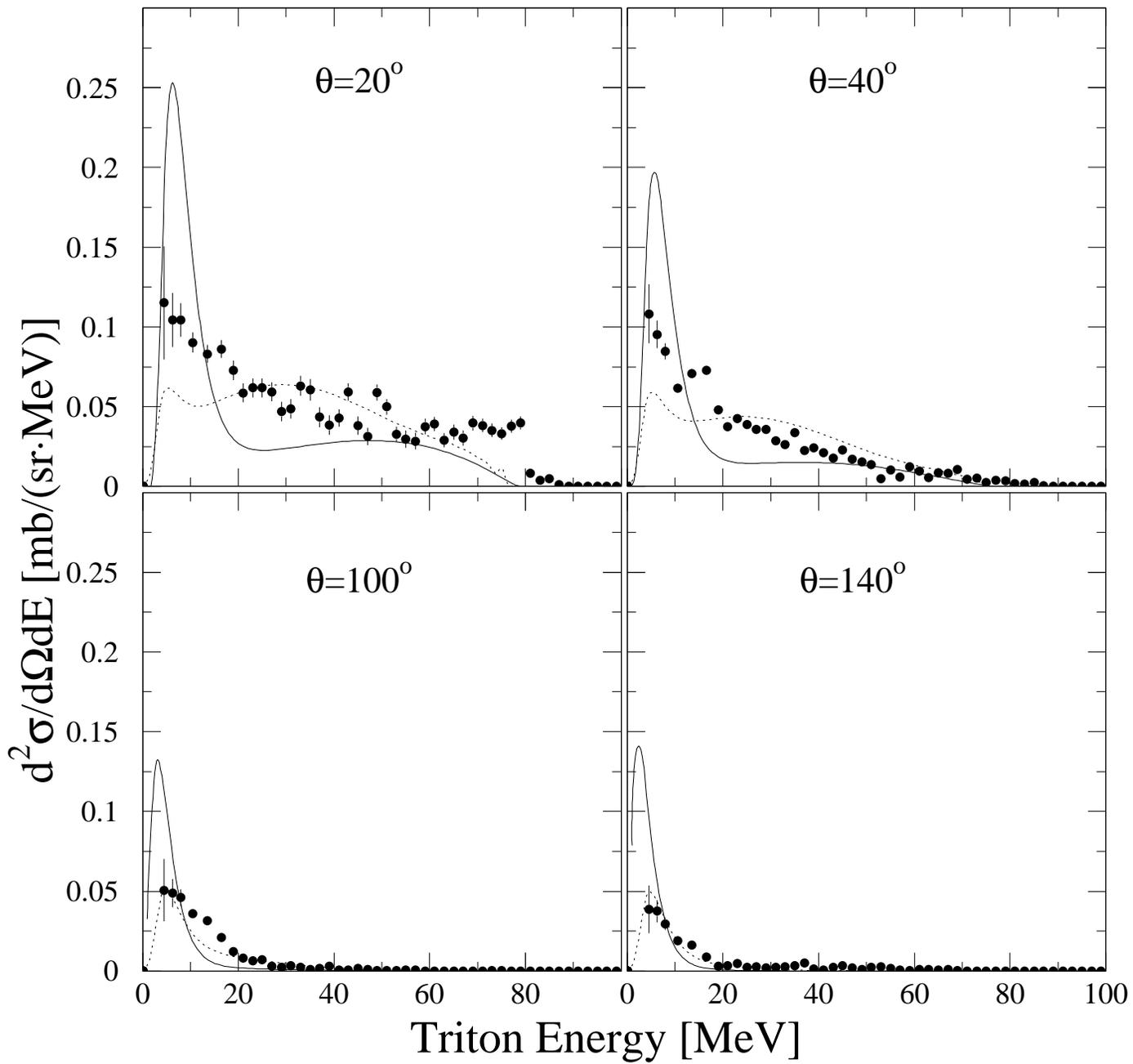
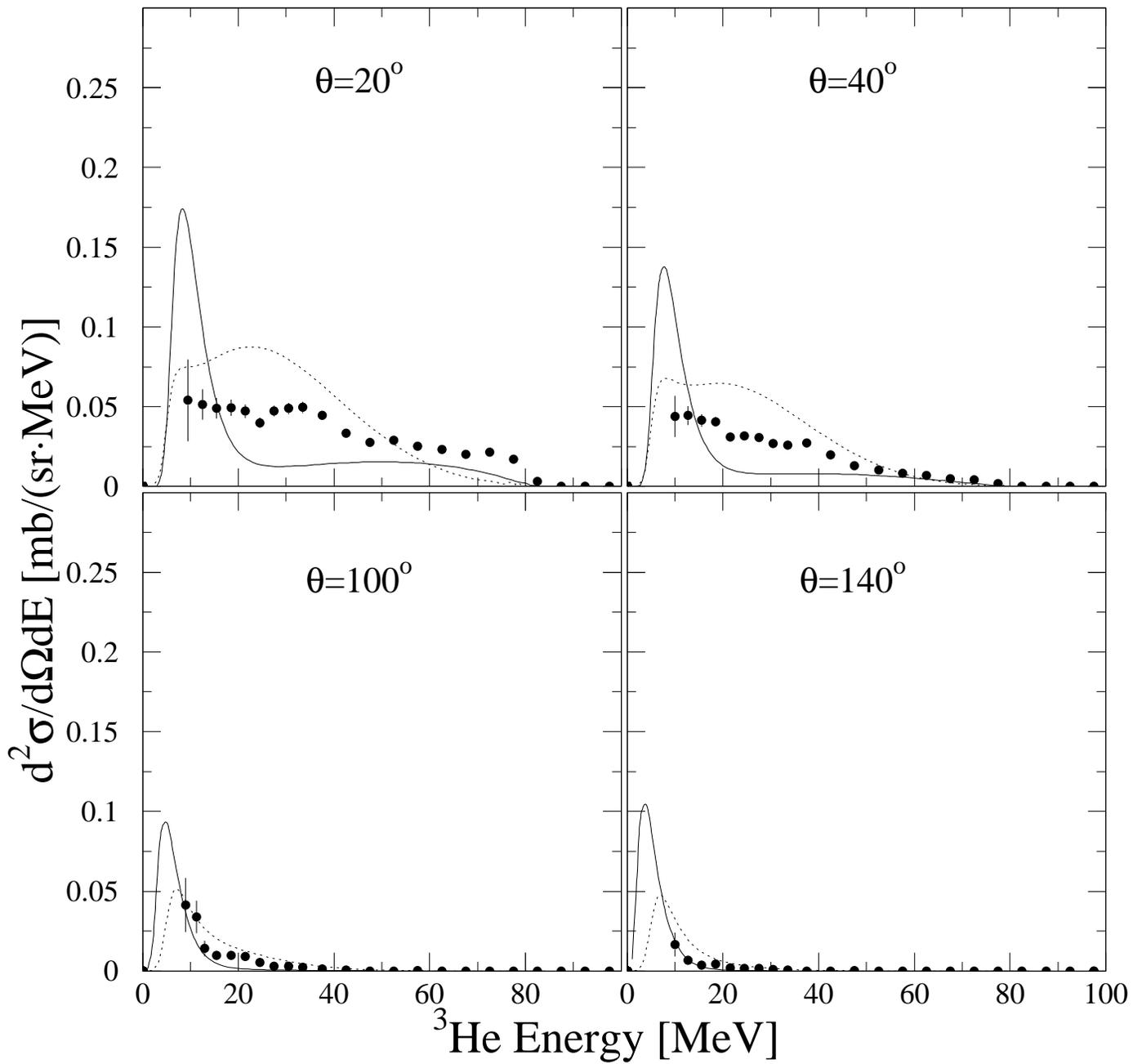
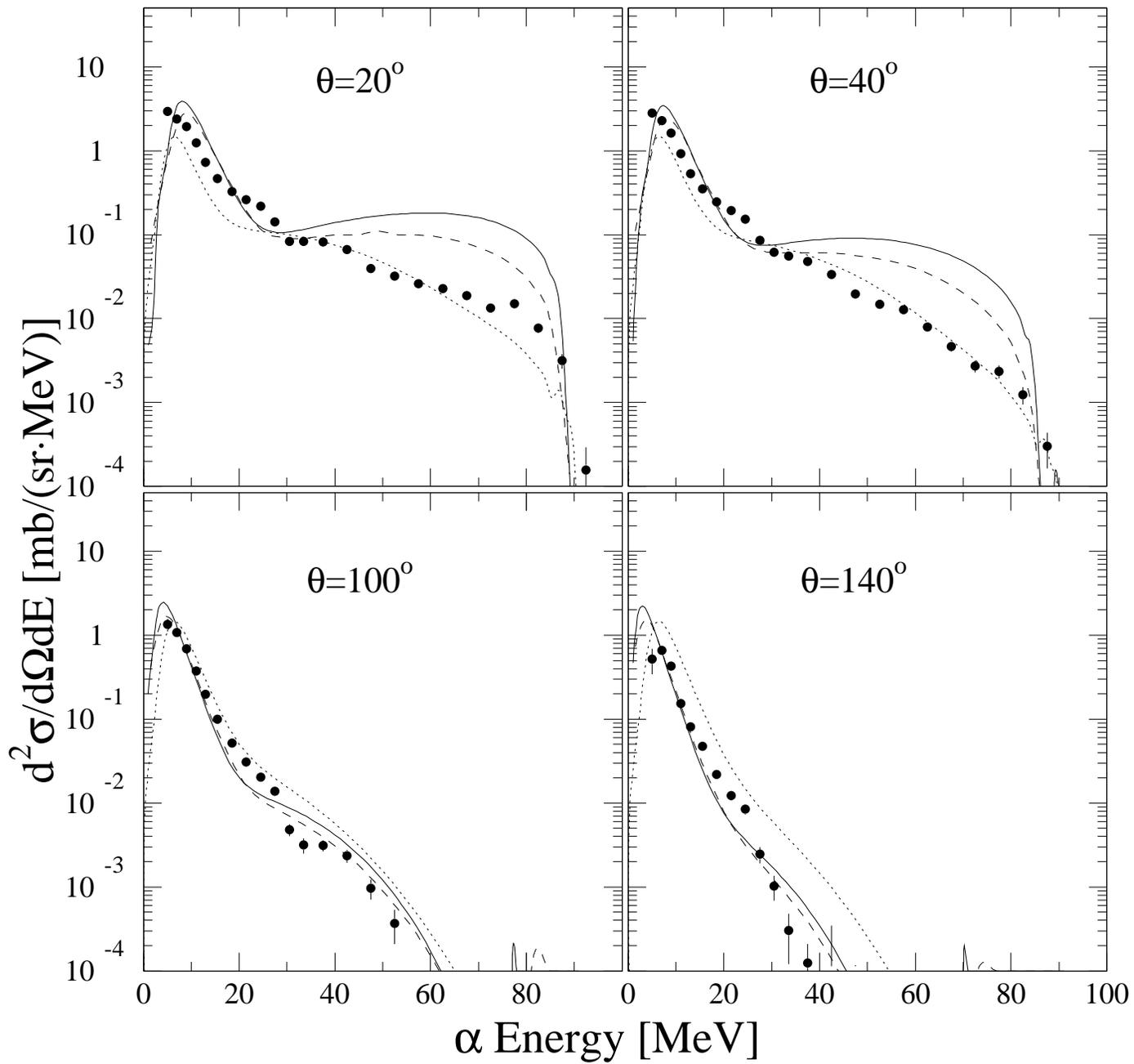
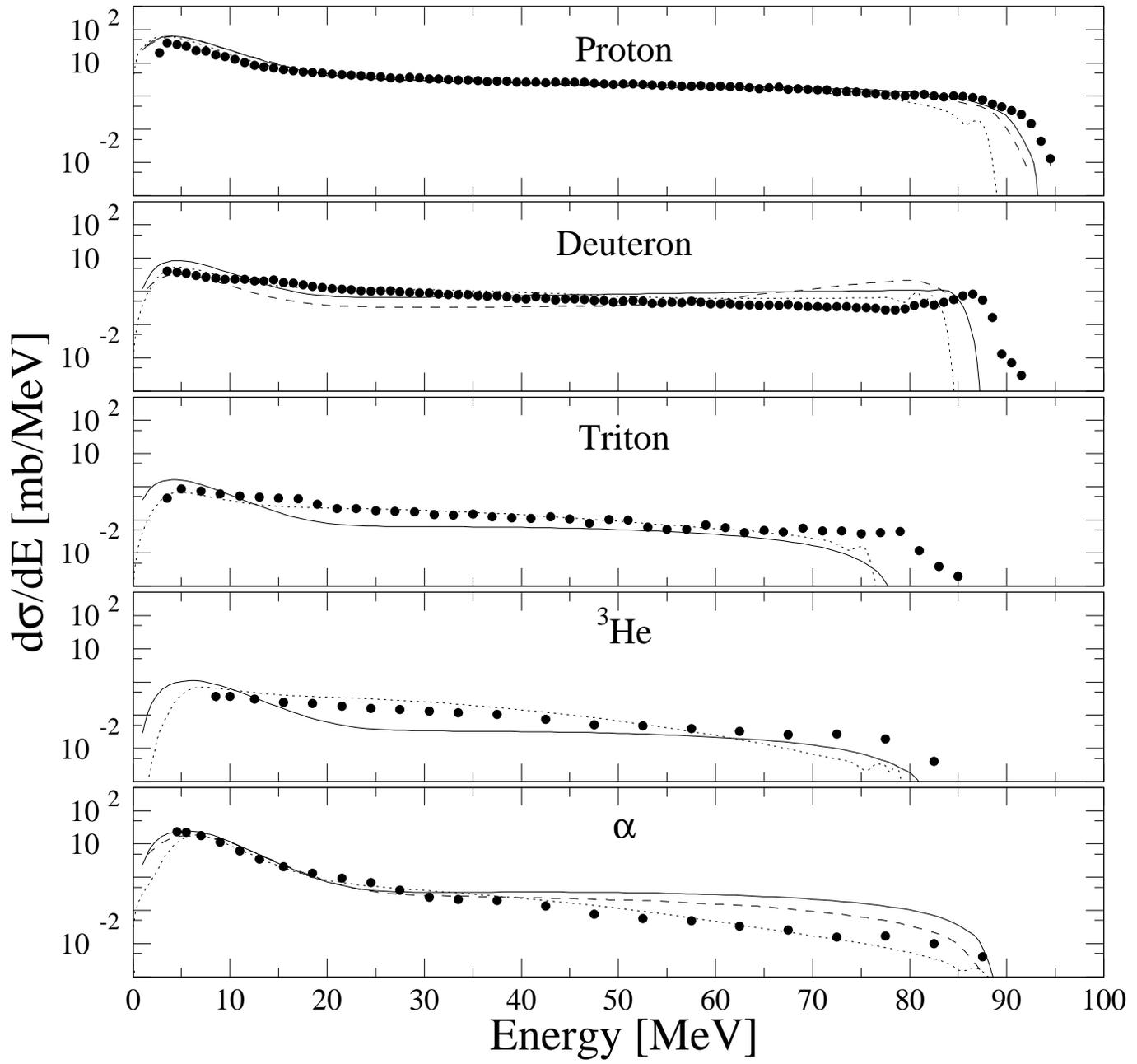
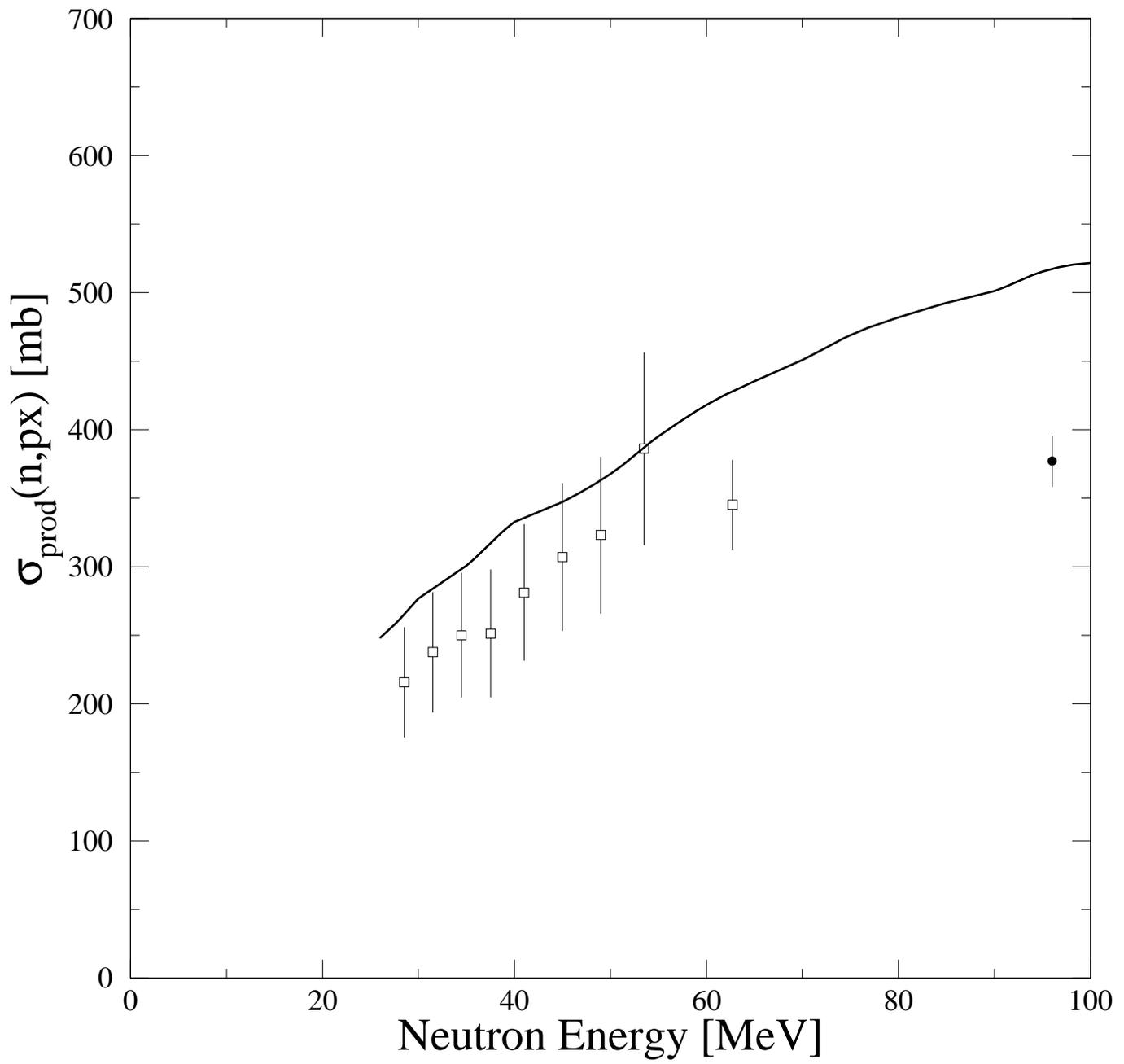
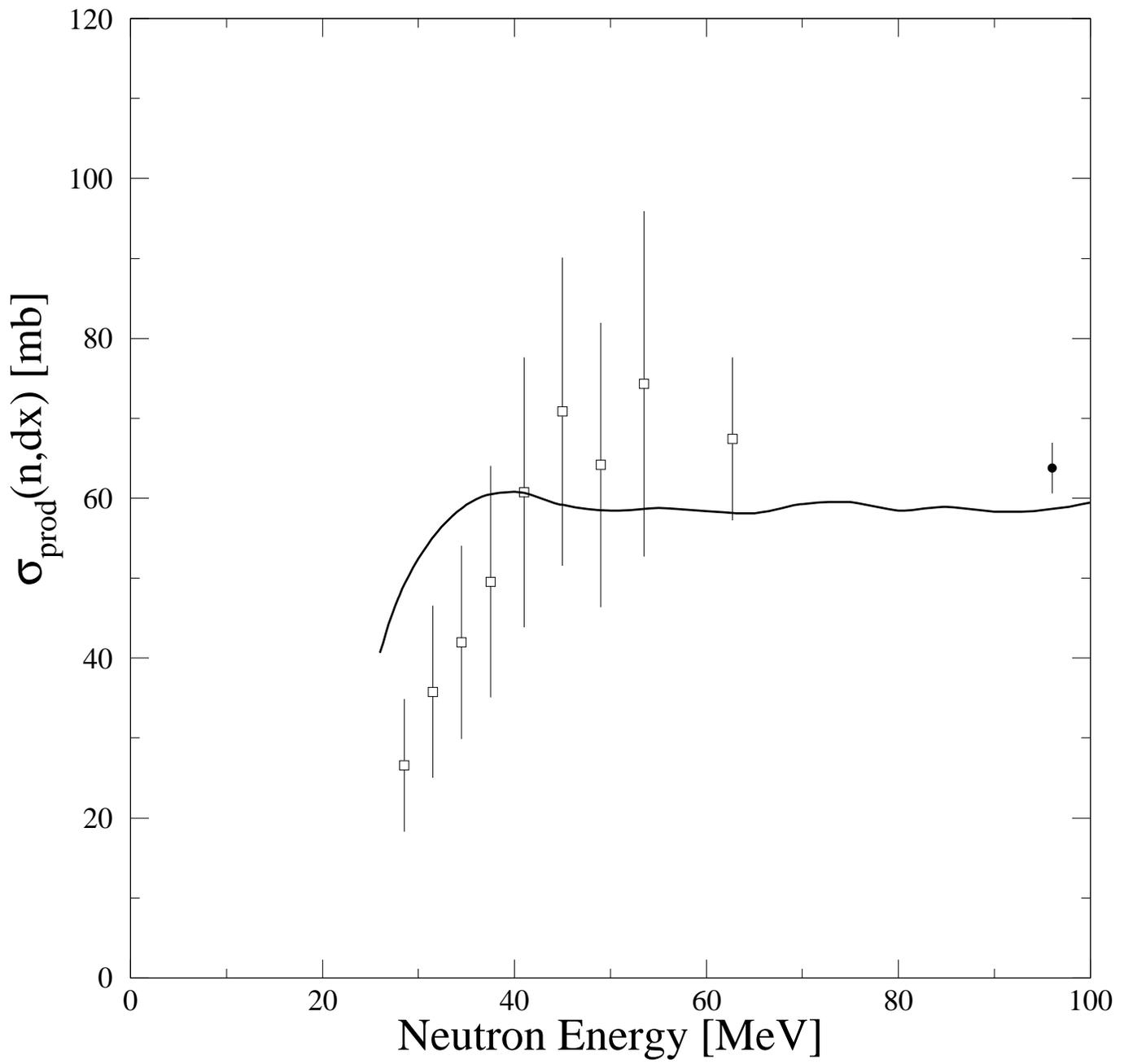
\end{document}